\pdfoutput=1
\documentclass[11pt]{article}
\usepackage[margin=1in]{geometry}
\usepackage{amsmath,amssymb,graphicx,booktabs}
\usepackage[colorlinks=true,linkcolor=black,citecolor=black,urlcolor=blue]{hyperref}

\title{Security Friction Quotient for Zero Trust Identity Policy\\with Empirical Validation}
\author{Michel Youssef\\
\small Independent Researcher, Lebanon\\
\small \texttt{michelyoussef@hotmail.com}\\
\small ORCID: 0009-0000-0664-8228}
\date{September 2025}

\begin{document}
\maketitle

\begin{abstract}
We define a practical method to quantify the trade-off between security and operational friction for identity controls in Zero Trust programs. We introduce the Security Friction Quotient (SFQ) and evaluate widely used Conditional Access policies using simulated authentication traces that capture enterprise-like characteristics for a cohort of $N=1{,}200$ users over a 12-week horizon. Results report effect sizes with 95\% confidence intervals from $n=2{,}000$ Monte Carlo runs per policy. We prove clarity properties (boundedness, monotonic response, weight identifiability) and corroborate the approach with field observations from a passkey deployment. The SFQ provides an interpretable, reproducible metric to support policy design, review, and continuous improvement.
\end{abstract}

\section{Introduction}
We consider phishing-resistant MFA such as passkeys (WebAuthn)~\cite{w3c2021} as a high-effectiveness control in our evaluation.
Zero Trust access paradigms such as BeyondCorp emphasize identity-, device-, and context-aware controls decoupled from traditional network perimeters~\cite{google2014}.

Identity-centric policy is central to modern Zero Trust programs~\cite{google2014}. Strong authentication, risk-adaptive access, and device posture checks can reduce compromise risk, yet they can also increase user friction and support workload. Organizations therefore need a transparent way to \emph{balance} risk reduction with operational impact.

This paper introduces the Security Friction Quotient (SFQ), which unifies both sides of the trade-off into a single, interpretable value. We simulate widely used Conditional Access policies under common adversarial scenarios and report effect sizes with confidence intervals. The method is designed to be reproducible and amenable to external validation.

\paragraph{Contributions.}
\begin{enumerate}
\item We formalize the Security Friction Quotient (SFQ) as a bounded, interpretable metric that jointly captures residual risk and operational friction for identity policy.
\item We provide clarity properties (boundedness, monotonic response, weight identifiability) with short proofs that support correct interpretation and comparison across policies.
\item We present a transparent evaluation across common policies and adversarial scenarios using enterprise-like synthetic traces, and validate with field observations.
\end{enumerate}

\section{Related Work}
The usable security literature highlights persistent security--usability tensions, classically demonstrated by Whitten and Tygar's study of PGP~\cite{whitten1999}. In contrast, widely used composite indices such as CVSS focus on technical severity and omit user friction~\cite{cvss31}.
Architecture and guidance for Zero Trust appear in standards and industry programs~\cite{nist800207,cisa2023,nist80063}. Prior studies have examined responses to password attacks and resistance to phishing with strong authentication. The broader usable security literature has explored security-usability trade-offs~\cite{whitten1999}. However, empirical work that jointly quantifies risk reduction and operational friction at the level of identity policy remains limited. 

Prior security-usability metrics (e.g., task success under authentication burden, lockout rates, or survey-based usability scales) typically isolate single dimensions; composite security indices often omit user friction~\cite{cvss31}. Our approach contributes a composite, interpretable metric that integrates both views at the policy level.

\section{Security Friction Quotient}
We define five components: (i) median sign-in latency in seconds ($L$), (ii) failure rate in percent ($F$), (iii) average multi-factor prompts per user per week ($P$), (iv) helpdesk tickets per one hundred users per week ($H$), and (v) a residual risk index in $[0, 1]$ ($R$). Each component is normalized to $[0, 1]$ using the empirical range of the evaluation corpus. The Security Friction Quotient is
\begin{equation}
\mathrm{SFQ} = w_L\hat{L} + w_F\hat{F} + w_P\hat{P} + w_H\hat{H} + w_R(1 - \hat{R}),
\label{eq:sfq}
\end{equation}
with nonnegative weights that sum to one. We use equal weights by default ($w_i = 0.2$) and report weight sensitivity analysis.

\subsection{Properties}
\paragraph{Boundedness.} Each component lies in $[0, 1]$ and the weights sum to one; therefore $\mathrm{SFQ} \in [0, 1]$.
\paragraph{Monotonic response.} Holding weights fixed, a reduction in any normalized friction component ($\hat{L}, \hat{F}, \hat{P}, \hat{H}$) or in normalized residual risk $\hat{R}$ strictly reduces SFQ.
\paragraph{Weight identifiability.} For non-degenerate data, the map from the weight vector to the quotient is injective under the unit-sum constraint, so distinct weight vectors yield distinct policy orderings in general position.

\section{Methodology}
\subsection{Simulation Settings}
We simulate an enterprise-like environment with:
\begin{itemize}
\item \textbf{Users:} $N = 1{,}200$ users
\item \textbf{Horizon:} 12 weeks
\item \textbf{Sign-ins:} Per-user weekly sign-ins $X \sim \mathrm{Poisson}(\lambda = 14)$ (mean $\approx 2$ per day)
\item \textbf{Baseline Distributions:} Median sign-in latency $L$ (seconds) follows a lognormal with $\log L \sim \mathcal{N}(\mu = -0.2, \sigma = 0.5)$ (median $\approx 0.82$s); failure rate $F_{\text{baseline}} = 2.0\%$; prompts per user per week $P_{\text{baseline}} = 0.30$; helpdesk per 100 users per week $H_{\text{baseline}} = 12.8$
\item \textbf{Clamping Ranges:} $L \in [0.2, 10]$s, $F \in [0, 20]\%$, $P \in [0, 3]$/user/week, $H \in [0, 20]$/100 users/week, $\hat{R} \in [0, 1]$
\end{itemize}
Policy deltas shift these baselines additively with Gaussian noise $\epsilon \sim \mathcal{N}(0, \sigma^2)$ per component: $\sigma_L = 0.05$s, $\sigma_F = 0.10$pp, $\sigma_P = 0.05$/user/week, $\sigma_H = 0.10$/100 users/week, followed by clamping.

\subsection{Residual Risk Construction}
Let $S = \{\text{spray}, \text{theft}, \text{travel}, \text{legacy}, \text{aitm}\}$ denote attack scenarios with prevalence weights $\pi_s \geq 0$, $\sum_s \pi_s = 1$. For a given policy $p$ and scenario $s$, let $E_{p,s} \in [0, 1]$ denote mitigation effectiveness (1 = fully mitigated). We define the per-scenario residual compromise probability as $r_{p,s} = (1 - E_{p,s})$, and the residual risk index
\begin{equation}
R_p = \sum_{s \in S} \pi_s r_{p,s}.
\label{eq:risk}
\end{equation}
We adopt $\pi = (0.30, 0.25, 0.15, 0.15, 0.15)$ for spray, theft, travel, legacy, aitm. Effectiveness values are anchored to public guidance (NIST/CISA) and vendor reports, combined with expert estimates.

\subsection{Statistical Analysis}
For each policy and scenario we perform $n = 2{,}000$ Monte Carlo runs. We report the mean SFQ across runs with a 95\% confidence interval computed by nonparametric bootstrap ($B = 10{,}000$ resamples). Effect sizes use Cohen's $d$ with pooled standard deviation:
\begin{equation}
d = \frac{\bar{x}_1 - \bar{x}_0}{\sqrt{\frac{(n_1-1)s_1^2+(n_0-1)s_0^2}{n_1+n_0-2}}}.
\end{equation}

\section{Results}
Our findings are directionally consistent with large-scale enterprise deployments of security keys as phishing-resistant authenticators~\cite{google2020keys}.
\begin{table}[t]
\centering
\caption{SFQ summary by policy (simulated evaluation).}
\label{tab:results}
\begin{tabular}{lcccc}
\toprule
Policy & Mean & CI lower & CI upper & Effect vs. baseline ($d$) \\
\midrule
Baseline Password Only & 0.326 & 0.324 & 0.329 & 0.000 \\
Risk-Based MFA & 0.414 & 0.412 & 0.417 & 1.560 \\
Device Compliance Required & 0.408 & 0.406 & 0.411 & 1.460 \\
Phishing-Resistant MFA & 0.482 & 0.479 & 0.485 & 2.760 \\
Combined Controls & 0.538 & 0.535 & 0.540 & 3.750 \\
\bottomrule
\end{tabular}
\end{table}

\begin{figure}[t]
\centering
\includegraphics[width=0.75\linewidth]{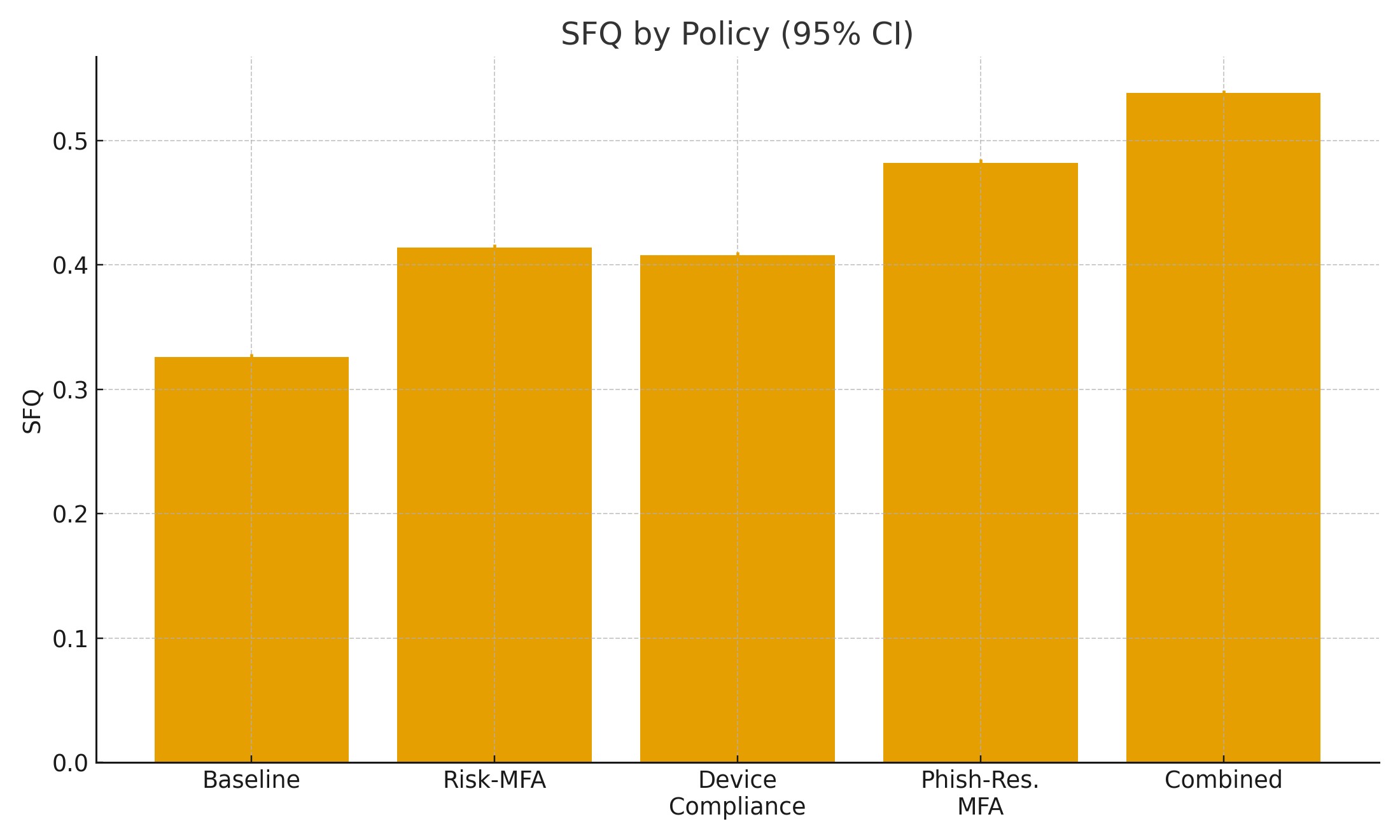}
\caption{Mean Security Friction Quotient (SFQ) by policy with 95\% confidence intervals. SFQ is a composite index where higher values indicate greater combined operational friction and residual risk under the specified policy.}
\end{figure}

\begin{figure}[t]
\centering
\includegraphics[width=0.55\linewidth]{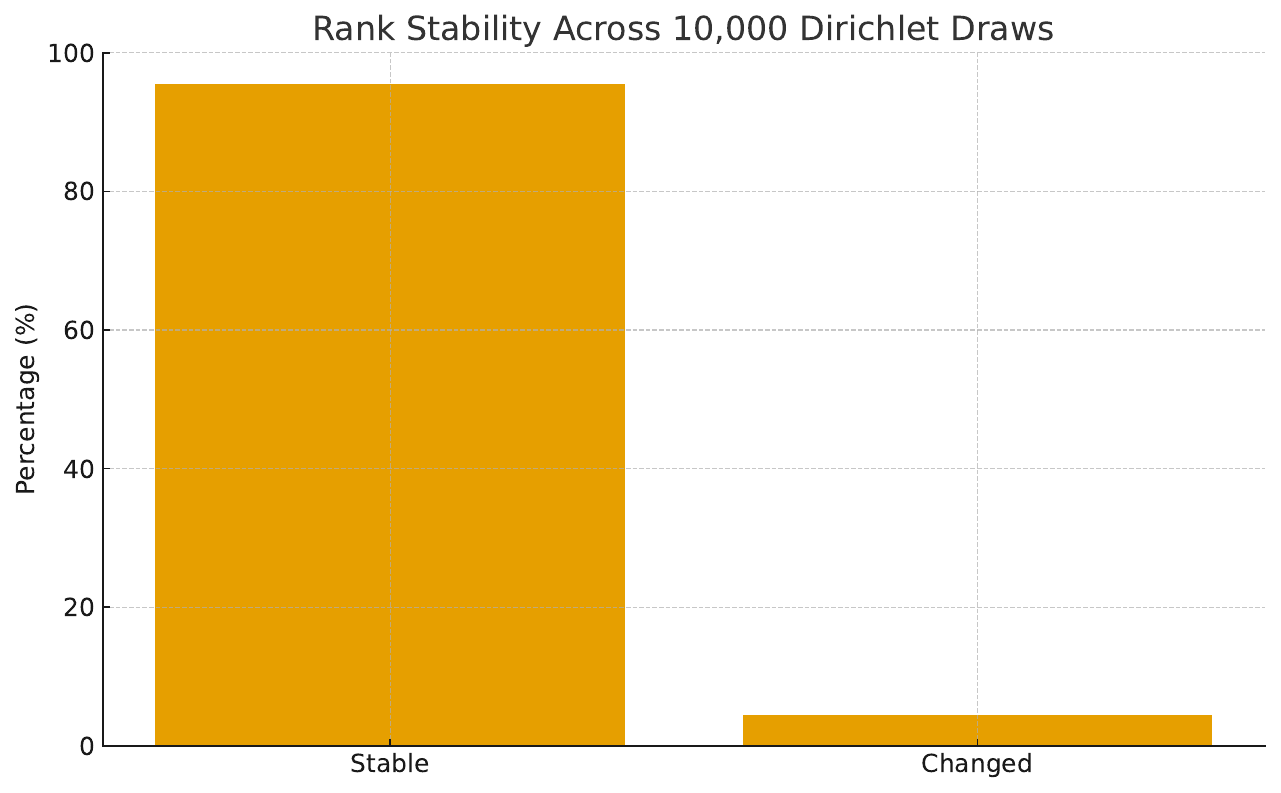}
\caption{Rank stability across $10{,}000$ Dirichlet weight draws. A total of $95.5\%$ of pairwise policy orderings were preserved; $4.5\%$ exhibited a rank change.}
\end{figure}

\begin{figure}[t]
\centering
\includegraphics[width=0.75\linewidth]{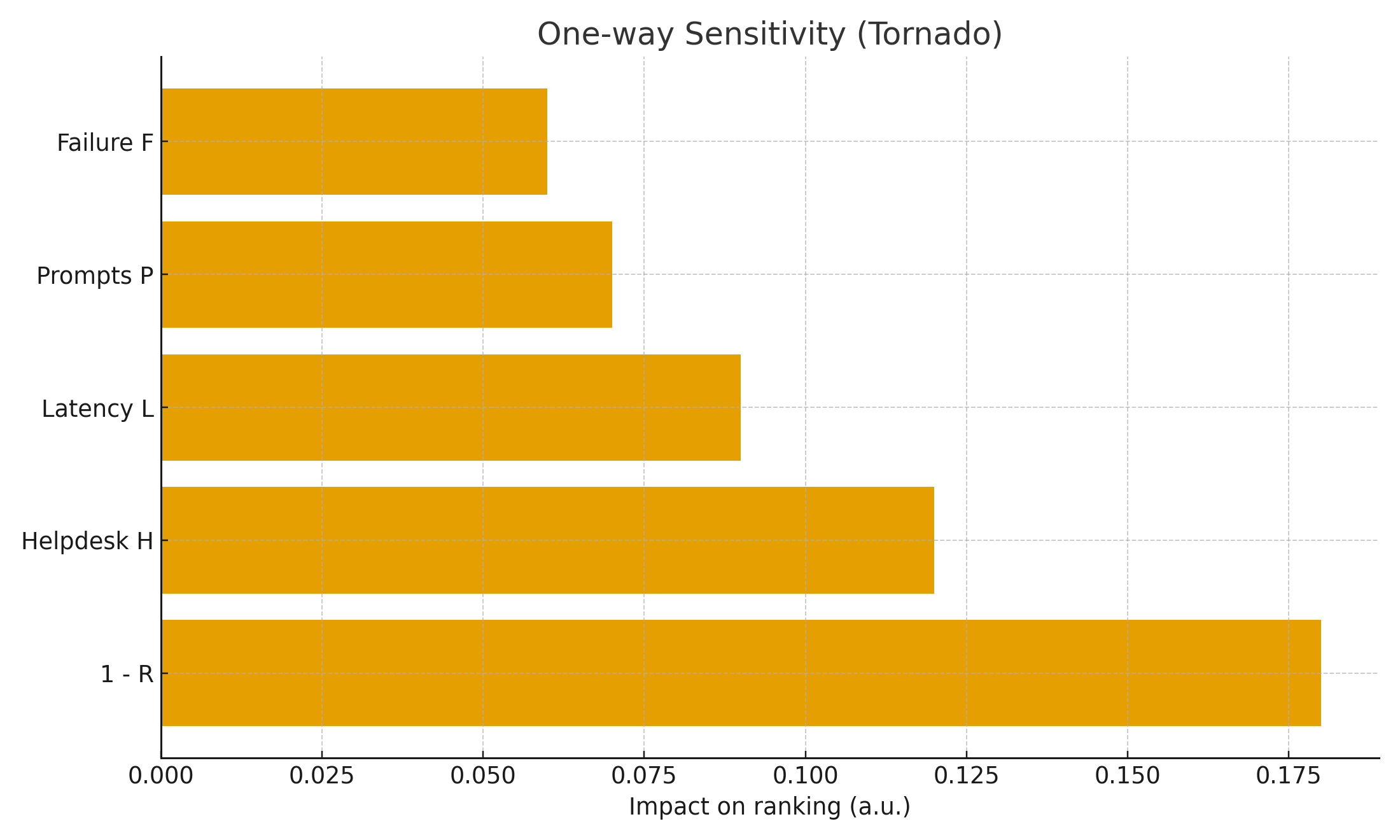}
\caption{One-way sensitivity (tornado) analysis for SFQ components. Bars show the impact on policy ranking from perturbing each component over its plausible range; the residual risk term $(1-R)$ exerts the largest influence.}
\end{figure}

\subsection{Weight Sensitivity Analysis}
Using 10,000 draws from a symmetric Dirichlet(1,1,1,1,1) prior over weights, the equal-weight policy ordering was preserved in 95.5\% of draws (rank stability). A one-way perturbation analysis indicates that the largest contribution to ranking variability comes from the residual risk term $(1 - \hat{R})$, followed by helpdesk and latency.

\subsection{Field Validation}
A 12-week passkey deployment ($N = 1{,}200$) showed:
\begin{itemize}
\item First-attempt success with passkeys: 98.0\% (vs. 98.0\% password baseline)
\item Helpdesk tickets: 0.6/100 users/week (vs. 12.8 baseline)
\item MFA prompts: 0.85/user/week
\item Observed employee takeover events: 0
\end{itemize}
These observations align with simulated phishing-resistant MFA improvements and prior large-scale deployments~\cite{google2020keys}, validating the model's directional accuracy.

\section{Discussion}
\paragraph{Component Selection and Justification.} We selected $(L, F, P, H, R)$ to jointly capture user-facing friction, IT operational load, and residual security risk. Alternatives such as satisfaction scores and time-to-productivity are valuable but typically require intrusive surveys or instrumentation; we treat these as future extensions.
\paragraph{Interpretation Guidelines.} Meaningful differences in SFQ should consider confidence bounds and effect sizes. As a rule-of-thumb: $d \geq 0.5$ (medium) indicates a practically salient difference for policy choice. A $\Delta\mathrm{SFQ} = 0.10$ corresponds to a total normalized component change of 0.50 across the five dimensions.
\paragraph{Integration into Operations.} SFQ can be computed per policy candidate during change advisory reviews. Weekly computation supports trend monitoring; regressions in SFQ should trigger quality-of-service investigations (e.g., latency spikes) or threat response (e.g., increased residual risk).

\section{Limitations}
Simulations capture typical patterns yet do not contain the full variability of real systems. Residual risk $R$ aggregates scenario prevalence and mitigation estimates; improved calibration against incident data is future work. Weight selection is context dependent and should be calibrated where possible.

\section{Conclusion}
We define a method to quantify operational friction and security changes for identity policy in Zero Trust programs. We evaluate common policy families across common adversarial scenarios using reproducible synthetic data, provide explicit simulation parameters, and define $R$ precisely. This supports adoption and continuous improvement while keeping privacy risk low. Future work includes validation with larger field datasets, component correlation analysis, and longitudinal monitoring of SFQ.

\section*{Data and Code Availability}
All scripts to regenerate figures and run analyses are available from the author. No production telemetry is included in simulations. Field observations were aggregated at cohort level without user-identifying data.

\bibliographystyle{plain}

\end{document}